# Impact of crystallinity on the circular and linear dichroism signals in chiral perovskite

*Reshna Shrestha[1], Wanyi Nie[1*]*

**Affiliations:**

[1]Department of Physics, University at Buffalo, State University of New York, Buffalo, New York 14260, USA

*Corresponding to: wanyinie@buffalo.edu

Funding: This work is supported by National Science Foundation DMR-2532768 and the research foundation of the state university of New York (RFSUNY) at the University at Buffalo.

Keywords: chiral perovskites, crystal orientation, circular dichroism, anisotropic absorption, optical artifacts

**Abstract**

Chiral perovskites owing to their broken mirror symmetry exhibit selective absorption of circularly polarized light manifesting a strong circular dichroism (CD). CD spectroscopy has been a key technique to understand chiral perovskites and how these semiconductors achieve chirality in molecular level. However, there is a debate on whether the observed CD is intrinsic to the chiral crystal structures or is modulated by extrinsic phenomena particularly linear dichroism (LD) and linear birefringence (LB) effects. This work investigates the chiroptical properties of (R-/S-MBA)$_2$CuCl$_4$ (MBA = Methylbenzyl ammonium) series by thoroughly studying the contribution from LD and LB to the observed CD signals. The comparison of highly oriented and randomly oriented films exhibit notable LD and LB contributions to the observed CD, which are caused by orientation-dependent electric-field interactions and local anisotropy. Both randomly and highly orientated films exhibit distinct CD responses, with LD–LB effects largely dominating the CD in highly oriented films. This has been revealed by the obvious shift in the observed CD signals to the below-absorption-edge (~ 430 nm) regime and broadening of features. Our findings demonstrate that careful consideration of crystal orientation and structural effects is necessary for appropriate interpretation of CD spectra in chiral perovskite thin films.

## 1. Introduction

Chirality has been a broad topic establishing its importance in variety of fields including physics, chemistry, biology, medical as well as pharmaceutical[1-5]. It is defined as a geometric property of an object not being able to superimpose with its mirror image, meaning the lack of mirror symmetry[1,6-8]. Leveraging the concept of chirality, semiconductors in chiral crystal structures exhibit distinct chiroptical properties, like circular dichroism (CD), defined by the unequal absorption of left-handed ($\sigma^+$) and right-handed ($\sigma^-$) circularly polarized light. In chiral systems, prefixes R and S are employed to define right- and left-handedness of the system respectively. The selective light-matter interaction in the chiral systems is attributed to the intrinsic coupling between electronic states and handedness, making them applicable in circular polarized photoluminescence[9-13], circular polarized detector[14-18], spin optoelectronics[19-28], direct polarization sensors[16,17,20,29-31], and quantum information[20,23,25-27,32-34].

Hybrid organic-inorganic perovskites are formed by inorganics co-crystallized with organic chiral linkers, that have been recognized as promising chiral semiconductors[26,30,35-38]. The incorporation of chiral organic ligands to the inorganic metal-halide frameworks results in the overall symmetry breaking from the achiral centrosymmetric forms into chiral space groups such as C2, and P2$_1$2$_1$2$_1$

– two typical space groups that adapt chiral symmetry[26]. The optical and electronic properties are directly tied to the tilting or distortion of the inorganic metal halides in the direction of intrinsic chirality of organics superimposing selectivity to circularly polarized light and spin polarized carriers due to the broken inversion and mirror symmetries[26,39-41].

CD spectroscopy that directly detects the difference of absorption of $\sigma^+$ and $\sigma^-$ photons, is a strong proof of the chiroptical properties of chiral perovskites. Depending on crystal dimensions, composition, and measurement geometry, reported CD signals for chiral perovskites range greatly, from 155 mdeg[35] to remarkably large values approaching 3200 mdeg[31]. Through superstructural chirality, larger CD value of 6350 mdeg has been reported in chiral perovskite metasurface[42]. According to Zhihang Guo et al.[43], giant optical activity showing molecular CD ~ 1800 mdeg has been observed in the 2D chiral perovskite of copper (Cu) co-crystallized with chiral R-/S Methylbenzyl ammonium (R-/S-MBA) in layered (R/S-MBA)$_2$CuCl$_4$ structure. While these significant CD signals reflect the intense chiroptical properties associated with the underlying chiral structures, there has been an active debate regarding the origin of CD responses through intrinsic chirality and external experimental artifacts[33,44-49]. The realization of the CD measurements can be strongly affected by linear dichroism (LD) - the differential absorption of orthogonal components of linearly polarized light. LD arises from the inherent anisotropy of layered crystal orientation in perovskites that generates imperfect polarization optics[50]. As a matter of fact, the crystal orientation in thin films promotes the complicated coupling of LD in CD signals along with linear birefringence (LB), an effect describing the differential transmission induced by different refractive indices of material[49-51]. Prevailed CD spectroscopy studies frequently seem to disregard LD and LB effects on true CD signals till recent works started to illustrate the complexity of the CD spectroscopy[44-49]. A recent work by Moon et al.[33] demonstrated that the LD artifacts arise from crystal orientation. Di Bari's earlier chiroptical studies demonstrated that certain organic thin films with macroscopic anisotropy exhibit abnormal CD responses that differ dramatically depending on the angle of incidence or the direction of light propagation[33,52,53]. The optical interference of thin film's LD and LB, also known as the LDLB effect, is directly linked to the observed abnormality in CD[33]. Inspired by these studies, in the Moon et al.'s work[33], it has been shown that the complexity of the CD signal can be smartly addressed by averaging observed CD signals coming from opposite crystal orientations. This was achieved by flipping the sample by 180° with respect to the light propagation axis. In the experimental regime, this method can be carried by taking the CD measurements through front and back surfaces reflection techniques, accounting LDLB effects. The genuineness of observed CD signals has been elusive. In Alexandre Abhervé's work[54], it has been showed that how the observed CD is in fact the combination of an actual CD, that can be obtained with previously mentioned front-back approach and LDLB response. The actual CD is intrinsic to chiral structure of the hybrid perovskite material while the LDLB response is due to inherent anisotropy of the crystal structure. The anisotropic alignment within the thin film together with possession of different refractive indices depending on the polarization of the light promote LDLB response[44,55]. These studies emphasize how crucial it is to separate orientation-induced contributions from the actual CD, which encourages a systematic examination of how crystal orientation shapes the observed chiroptical response.

In our work, we investigate a (R-/S-MBA)$_2$CuCl$_4$ (R-Cl and S-Cl hereafter) compound with controlled crystal orientations produced by slow-diffusion method to elucidate the LD and CD properties. We quantify LDLB artifacts in CD measurements based on these contrasting orientations. Firstly, we examined the CD utilizing photo-elastic modulator-based scheme. In the observed CD signals, we found a significant LD contribution, which showed up as a baseline shift of around 60% and a clear LDLB feature at about 2.88 eV (~ 430 nm), which is significantly below the bandgap ($E_g \approx 3.26$ eV). These characteristics verified that the apparent CD in the low-energy region is dominated by LDLB. Inspired by the work done by Narushima and Okamoto[56], which enabled the discrete generation of circularly polarized light, we isolated the intrinsic chiral response using a beam-displacer-based method. This configuration effectively removed LDLB contributions and revealed the true CD signature (~ 380 nm) of the chiral perovskite.

Thus, with the systematic separation of LD signals from the observed CD due to the influence of crystal orientation, we determine how LD contributions couple into CD measurements and attempt to establish a correlation between structural anisotropy and polarization-sensitive absorption by contrasting polycrystalline films with well orientated single-domain crystals. Distinct measurements techniques applied in this work allows to garner understanding of chiroptical responses and their cause, consequently suppressing orientational artifacts and separating intrinsic CD response. Our work provides both fundamental understanding of light-matter interactions in chiral semiconductors and useful directions for accurate chiroptical characterization in device-relevant geometries by establishing a rigorous framework to differentiate true CD from extrinsic LD effects.

## 2. Results

### 2.1. Orientational and Structural Analysis of (R-/S-MBA)₂CuCl₄ Thin Films

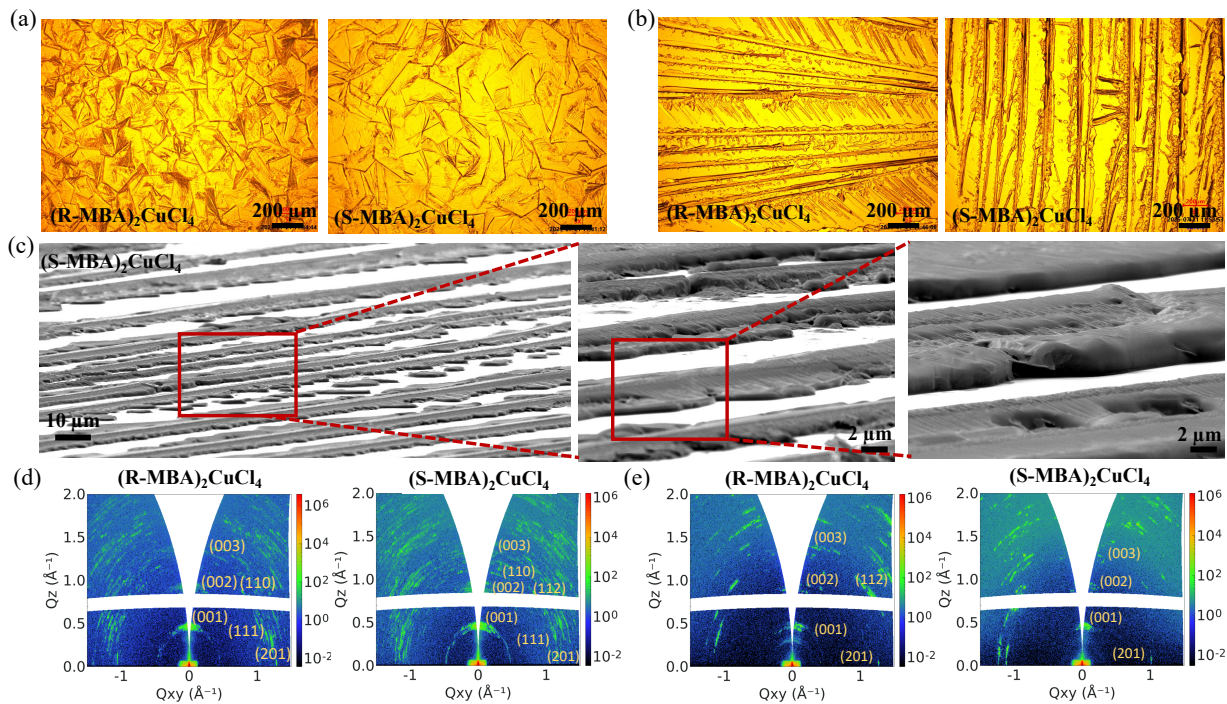

**Figure 1. Crystal structure and orientation analysis of (R-/S-MBA)₂CuCl₄ polycrystalline thin films.** Optical microscope images of (a) Typical spin-coated thin films and (b) Typical samples made with the slow diffusion method. (c) SEM images of a (S-MBA)₂CuCl₄ thin film made from the slow diffusion method. The GIWAXS pattern of (d) Spin-coated thin films (e) Slow-diffused thin films.

We have utilized R-/S-Methylbenzyl ammonium chloride (R-/S-MBACl) as a chiral linker to co-crystallize with metal halide framework, copper chloride (CuCl$_2$). The resulting R-/S-MBA$_2$CuCl$_4$ is a 2-dimensional, layered structure which follows a non-centrosymmetric monoclinic crystal structure belonging to chiral C2 space group[57,58]. We first focused on understanding how crystal orientation affects chiroptical properties. To control the crystal orientation, we adapted a slow solvent diffusion method that uses the concentration gradient to guide the crystal growth direction.

Briefly, we dropped liquid precursor on the edge of glass substrate and pulling the liquid along the other edge introducing solution gradient. Solvent will evaporate along the gradient where crystal will grow following the same direction. We first used optical microscopy (OM) images (Figure 1a-b) to confirm the crystal morphology fabricated by the slow diffusion methods in comparison with the spin-coated samples. According to the OM images, the slow-diffused samples (Figure 1b) exhibited periodic and prolonged dendrites like structures, that orient along a preferred direction spanning approximately 200 μm. In contrast, the spin-coated films displayed in Figure 1a are composed of micron-sized random domains structured without long-range order. To further visualize the crystal morphology in each dendrite, we employed the scanning electron microscopy (SEM) for verifying the micro-structures. Figure 1c depicts a typical SEM image for a slow diffused S-Cl thin film viewed from the side, which demonstrated well-stacked, layered crystallites with smooth borders. In the zoomed-in SEM image, we observed a closely packed layered structures in one dendrite, indicative of a highly oriented structure. Such a distinct feature on a micro-scale account for the lamellar arrangements of the perovskite structure which comprises parallel stacks organization exhibiting peculiar material specific multi-layered characteristics. The spin-coated films have a more isotropic and uneven morphology (SI Figure 01). The R-Cl sample showed similar features under SEM and are illustrated in SI Figure 02. As supported by both OM and SEM images, the crystal structures of the slow-diffused samples exhibit two distinct observable features: stripes (main extended backbones) propagating across certain region and branches that grow through the stripes. This pattern is presumably a result of the Mullins-Sekerka instability effect[59], which is a diffusion-limited morphological instability that results in non-uniform, branching, or dendritic growth patterns during crystal growth from solution. This instability is favored as a consequence of tiny disturbances at a moving solid-liquid interface increases rather than smoothing out[60]. In the context of, solution processed thin films, the solid-liquid interface establishes the dynamic junction between the expanding crystalline solid film and the surrounding precursor solution. This results in the transport of molecular or ionic species from the liquid phase into the solid lattice[61]. Whether the film develops morphological instabilities like branching or dendrites or grows smoothly depends on its stability. While an unstable interface magnifies disturbances and results in dendritic or branched morphologies, a stable interface smoothens out.

Grazing-incidence wide-angle X-ray scattering (GIWAXS) was utilized to understand the crystallographic orientations. Figure 1 d-e shows the GIWAXS patterns of thin films for both spin-coated and slow-diffused samples respectively. The peaks are indexed according to the simulated powder X-ray diffraction pattern (shown in SI Figure 03) from the single crystal structure obtained from the work reported by Lorenzo Malavasi et al.[58]. In the spin-coated films, we observed primarily broad diffraction rings and arcs that confirm randomly oriented grains in the films. The indistinguishable scattered spots and arcs in the spin coated thin films at higher q values are assumed to arise from the out-of-plane scattering from the randomly oriented crystalline surface. The intensified patterns within the lower $q_z$ values validated the 2D structure of the crystal. Although random, the growth patterns favored (001) crystallographic plane corresponding to ~ 0.5 Å$^{-1}$ $q_z$. The GIWAXS intensity mapping in the slow-diffused samples depicted cleaner patterns with reduced redundant features. The broadening of the scattering features was clearly reduced in the slow-diffused counterparts since, they are fabricated with controlled crystallization over a prolonged growth time. Intense scattering from the (001) planes were observed that were primarily along the out-of-plane direction. (201) plane was found to be along the in-plane direction. The crystallographic information collected through the GIWAXS closely relates to the reported results for single crystals[18]. We visualized (001) and (201) crystallographic planes in the VESTA software by utilizing the CIF (see SI Figure 04). Notably, both R-Cl and S-Cl preserve similar structural ordering, demonstrating that lattice alignment is unaffected by the organic spacer's chirality except for compound influenced patterns. This controlled crystal orientational anisotropy between spin-coated and slow-diffused samples is an explicit approach to distinguish the impact of crystal orientation on CD and LD signals.

## 2.2. Polarization-Dependent Linear Optical Anisotropy

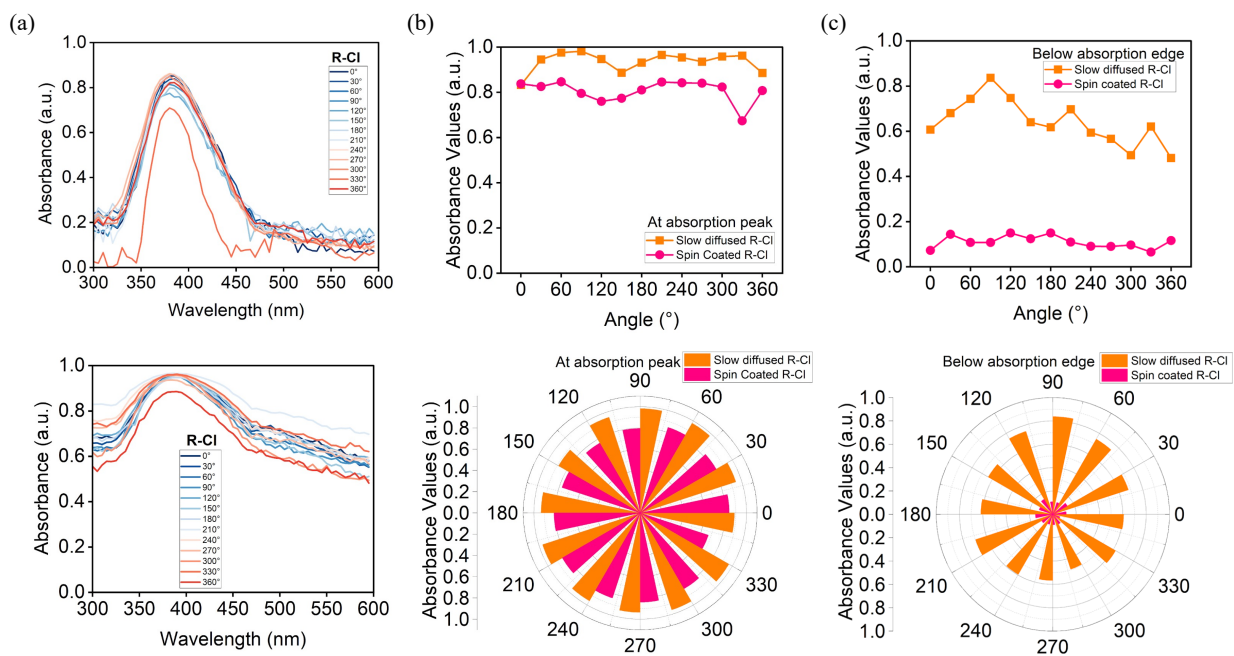

**Figure 2. Absorption spectra of (R-MBA)$_2$CuCl$_4$ thin films taken with linearly polarized light by tuning polarizer angles.** Absorption spectra of (a) Spin-coated (top) and slow-diffused (bottom) polycrystalline thin films. (b) Absorbance values at the absorption peak position (~ 380 nm) and (c) Absorbance values below the absorption edge plotted as a function of linear polarizer orientation angle relative to the horizontal line in our lab reference frame.

Next, we studied the optical absorption spectra of both spin-coated and slow-diffused R-/S-Cl thin films under linearly polarized light. We anticipate that LD effect will amplify in the highly oriented samples because the electrical field of the linearly polarized light interacts with transition dipoles aligning in the same direction and thus exhibits anisotropic absorption. We acquired a series of absorption spectra taken under linearly polarized monochromate light while varying the polarization axis' angles over a complete 0° to 360° rotation cycle with respect to the horizontal plane of the lab's reference frame. Figure 2a illustrates the linearly polarized absorption spectra for the spin-coated (top plot) and the slow-diffused (bottom plot) R-Cl thin film. We measured the absorption spectra by varying incident light polarization angles from 0° to 360° in a step of 30°. We observed an absorption peak located at ~ 380 nm that is attributed to Cu d-orbital (VB) to Cl s-orbital (CB) transition, which reproduces the previously reported absorption spectrum[43,57].

From the absorption spectra shown in Figure 2a, we observed that for different light polarization angles, there is a vertical position offset in the absorption peak in both spin-coated and slow-diffused R-Cl thin films while the peak position remained the same. We observed baseline shifting in the transmission region which overall affected the absorption spectra suggesting the presence of relatively significant LD signals on slow-diffused samples over spin-coated samples. The underlying optical anisotropy of the films is reflected in the absorption spectra, where the baseline offset and peak amplitude both clearly depend on the polarizer angle.

The amplitude of the absorption peak varied consistently with rotation of the polarizer; the peak appeared most noticeable when the electric field of the incident light coincides with the molecular transition dipole moment. This result demonstrated that polarization-dependent transition probabilities i.e., LD brought on by the transition dipoles' preferred orientation with respect to the crystal lattice, are the source of the absorption peak modulation. The absorption peak in spin-coated films, where crystal orientations are mostly random, fluctuated substantially with polarizer angle. This suggests an optical response that is almost isotropic. The angular dependency of peak amplitude is evident in slow-diffused films. This indicates a higher degree of crystalline or molecular alignment. The baseline shifts observed in the absorption spectra suggested light scattering that depends on the polarization angle of the incident light. We have summarized that the absorption peak values for both spin-coated, and slow-diffused R-Cl thin films remained nearly constant as shown in figure 2 (b). At the absorption peak, the fundamental electronic transitions are mostly comparable to the LD contributions in both spin-coated and slow-diffused thin films relative to light-polarization angles. The spin-coated thin film showed least variation in the peak values as a function of polarization angles. Although, the slow-diffused thin film followed similar trend, the peak absorbance values appeared at slightly higher positions attributed to baseline shifting.

To validate our observation, we plotted the absorbance values below the absorption edge, meaning in the transmission region (~ 590 nm), against the polarization angles as illustrated in figure 2 (c). As opposed to the absorption peak, the baseline offset below the absorption edge changed with polarization in a different way. After the absorption edge (off-peak area), this baseline shift became more noticeable, and its angular dependency indicated contributions from anisotropic light scattering or reflection rather than electronic processes. Measurable baseline shifts resulted from the anisotropically oriented microcrystals' distinct scattering or rerouting of light when the polarizer lines up with particular crystallographic orientations in the slow-diffused films. These scattering-induced effects are isotropic in spin-coated films by random crystal formation, resulting in a baseline that is almost flat and polarization-independent. Consequently, polarization-dependent scattering/reflection processes brought on by structural anisotropy in the film morphology are the main source of the baseline offset, whilst polarization-selective electronic transitions control the modulation of the absorption peak. These findings demonstrated the LD and associated optical anisotropies are induced by the larger degree of crystalline ordering in the slow diffusion grown samples. In the S-Cl thin films, we observed a similar pattern for both spin-coated and slow-diffused thin films. SI Figure 05 provides a thorough representation of the linear absorption spectra for S-Cl with varying incident light polarization angles.

*2.2.1. Linear Dichroism Insights*

The presence of LD can be correlated to the crystal orientation dependence of chiroptical signals and will be pronounced when there is robust anisotropy of material's optical transition dipoles. In the context of spin-coated thin films, LD contributions seem to be averaged out, while for oriented slow-diffused thin films having certain single-domain crystal features, LD signatures remain significant, thus giving a way for relevant interpretation of CD. For both spin-coated and slow-diffused thin films, the absorption peak position remained in the expected region ~ 380 nm. However, in the off-peak regime, below the absorption edge, we observed a strong difference in the spin-coated and slow-diffused thin films accompanied by the baseline shifting. If the electric field of the polarized light matches the orientation of the molecules in thin films, this will favor LD and we will observe shifting from the baseline, otherwise LD will have minimal contribution. By observing the absorption spectra, the contributions of LD in both the spectral peak region and the baseline region were distinguishable. In the absorption peak, the observed LD implies for the preferential polarization-specific absorption of light, whereas the LD contribution to the baseline is ascribed to polarization-specific scattering or reflection processes. The spin-coated thin films show random crystal growth exhibiting anisotropy, resulting in the cancellation of LD contribution probably due to equal and opposite LD values distributed across the thin film structures with almost isotropic absorption spectra for distinct polarization angles of incident light. However, for the slow-diffused counterpart, the tuning of polarizer angles significantly affected the baseline shift, implying how the crystal growth orientation has crucial role in optical properties.

## 2.3. CD (1f) and LD (2f) Harmonic Analysis

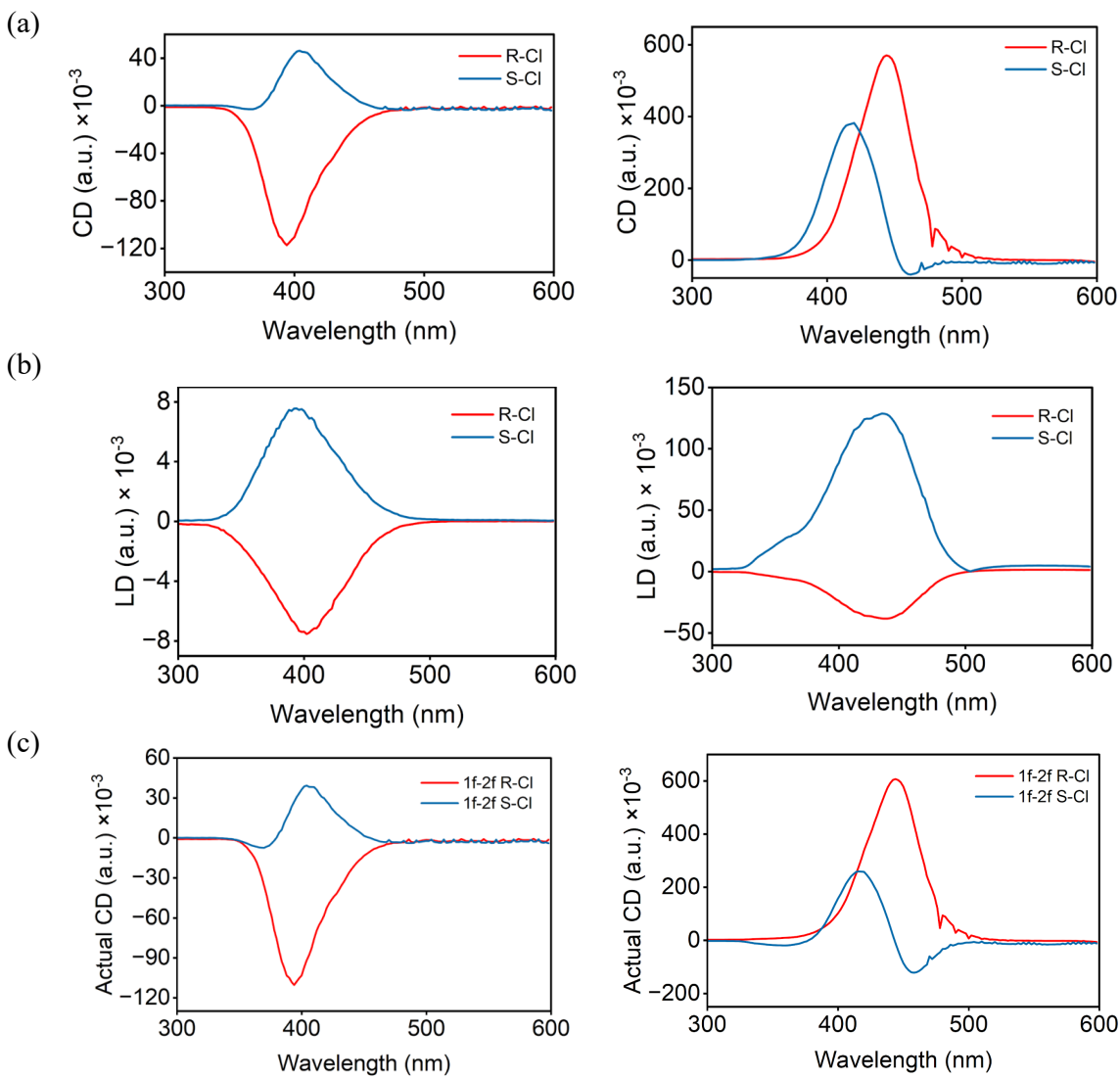

**Figure 3. Harmonic analysis to detect LD (2f) mixing in CD (1f) in (R-/S-MBA)$_2$CuCl$_4$ thin films**. (a) CD spectra. (b) LD spectra. (c) Actual CD spectra by (1f-2f) analysis. Here, the left

column and the right column of the figure comprise the spectra for spin-coated and slow-diffused thin films respectively.

To quantify our theory of the mixing of LD in the observed CD signals, we investigated the chiroptical responses of chiral perovskite thin films in two frequency modulations. The CD spectra and the LD spectra were measured at the modulator frequency (f) (with circularly polarized signals) and twice the modulator frequency (2f) (with linearly polarized signal) respectively. We utilized our in-house experimental CD system. The arrangement was based on the conceptual principle of commercial high-throughput CD spectroscopy, which uses a photo-elastic modulator to add a phase delay in one of the linearly polarized lights at a quarter of its wavelength in order to produce ($\sigma^+/\sigma^-$) circularly polarized light. The schematic utilized for this procedure is illustrated in SI Figure 06.

The resulting CD spectra of spin-coated (left) and slow-diffused (right) R-Cl and S-Cl thin films using the photo-elastic modulator-based approach are shown in Figure 3a. Near their absorption peak, at about 380 - 400 nm, we observed significant peaks for R-Cl and S-Cl spin-coated thin films corresponding to CD detected near the absorption band edge. The CD signals for left- and right-handedness appeared as mirror images to each other. Such features are consistent with previously published peak position in the as acquired CD spectra[43]. Below the absorption edge, the CD spectra exhibited tail-like features.

In the case of slow-diffused thin films, more intensified peaks were observed in the below-absorption band-edge regime. However, the CD signals did not show mirror images for distinct handedness. For S-Cl, although the peak was observed ~ 400 nm, additional features also appeared below the absorption edge. For R-Cl, the peak position shifted below the absorption edge further accompanied by multiple bumps. In principle, the intrinsic CD signals are expected in the absorption edge with a differential-like shape. However, the appearance of additional intense signals beyond absorption edge made us to expand our study beyond the CD and account for the possible contributions from the LD associated to the high crystallinity and orientation of crystal domains in the thin films.

To check the presence of LD effect in thin films, we conducted LD experiment at twice the modulator frequency (2f). As illustrated in figure 3b, the LD spectra for both spin-coated (left) and slow-diffused thin films (right) appeared in the absorption band-edge regime (~ 380 - 450 nm), confirming the LD contamination in the observed CD. As expected, the LD effect in slow-diffused thin films appeared almost two-order-of magnitude higher than that in spin-coated thin films. The mixing of LD signals in the actual CD signals may intensify the observed CD signals, especially in the slow-diffused thin films. The high crystallinity introduces the local anisotropy and the non-alignment of LD and LB transition dipoles making such kind of crystalline thin films highly dependent on the orientation and sample preparation methods[53,54]. As the samples exhibited significant LD responses equivalent to observed CD responses, we attempted to eliminate the LD artifact by performing the difference between CD (1f) and LD (2f) signals.

Figure 3c shows the quantification of CD-LD mixing by harmonic analysis. With the exception of the noticeable drop in signal amplitude, the actual CD responses showed a direct resemblance to the chiroptical properties as displayed on the observed CD as shown in figure 3a. The mixing of LD to the observed CD comes from the crystal orientation related imperfections such as misalignment and birefringence. Due to strong crystallographic orientations in the thin films, the complete elimination of LD effect from the intrinsic CD with harmonic discrimination between CD and LD signals was not an ideal approach. In principle, this approach should have thoroughly been able to isolate the LD contribution. The birefringence nature of the material could be responsible for incomplete separation. Thus, it is crucial to eliminate the LDLB effect from the observed CD, which will be discussed in subsequent module.

## 2.4. Chiroptical Characterizations: Isolation of Intrinsic CD via Front-Back Illumination

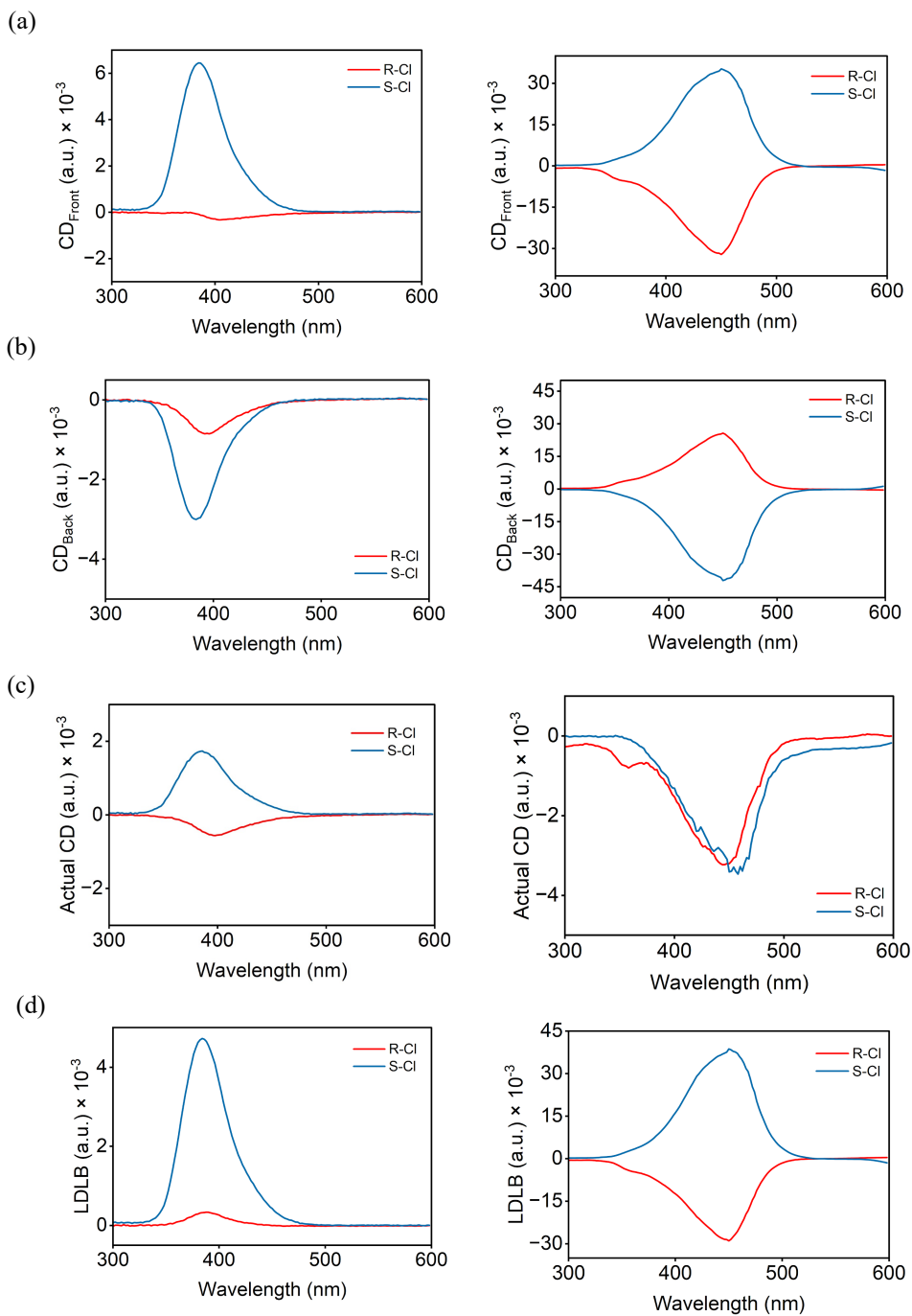

**Figure 4. Chiroptical characterization of R-/S-MBA$_2$CuCl$_4$ thin films with home-built CD setup.** (a) CD spectra taken from front illumination (CD$_{Front}$). (b) CD spectra taken from back illumination (CD$_{Back}$). (c) Actual CD spectra with semi-sum of front/back illuminations. (d) LDLB

spectra with semi-difference of front/back illuminations. Here, the left column and the right column of the figure comprise the spectra for spin-coated and slow-diffused thin films respectively.

Next, we adapted previously reported approaches for the careful analysis of chiroptical responses of chiral perovskite thin films. In Alexandre Abhervé's work[54], it has been highlighted on the non-negligible effect of LDLB in the actual CD. When projected in the plane orthogonal to the direction of light transmission, this designated LDLB phenomenon entails pairs of non-parallel transition dipoles. To better understand the intrinsic chiroptical properties, it is crucial to separate the actual CD from the LDLB contributions. In an effort to separate CD from LDLB, the thin film's chiroptical response was measured in two alternative setups: with the film facing the light source (front measurements) or facing the detector (back measurements)[33,54]. This was achieved by flipping the sample 180° with respect to the incident light transmission axis.

The actual CD response would be the semi-sum of the CD taken from the front and the CD taken from the back.

$$CD_{Actual} = \frac{1}{2}(CD_{Front} + CD_{Back}) \qquad (1)$$

When we flip the sample by 180° relative to the axis of light propagation, the contribution from the LDLB effect is inverted. As consequently, it was characterized as antisymmetric with regard to the direction of light propagation. To quantify the LD and LB artifacts, we plotted the LDLB signals by considering the semi-difference of front and back CD measurements.

$$LDLB = \frac{1}{2}(CD_{Front} - CD_{Back}) \qquad (2)$$

Figure 4(a-d) shows the CD spectra collected through the front illumination of light (CD$_{Front}$), the CD spectra collected through the back illumination of light (CD$_{Back}$), the actual CD spectra

demonstrated by performing the semi-sum of the CD front/back illuminations, and the LDLB spectra demonstrated by performing the semi-difference of the CD front/back illuminations for the both spin-coated (left) and slow-diffused (right) thin films respectively.

As for the spin-coated thin films, the $CD_{Front}$ signals were observed ~ 400 nm (Fig 4a). The peak intensity for S-Cl thin film was much intensified than that for R-Cl thin film, meaning suppressed CD. Then, the $CD_{Back}$ signals for the spin-coated thin films were detected in the similar region as the $CD_{Front}$ signals, ~ 400 nm (Fig 4b). In the back measurement, the CD responses for both R-Cl and S-Cl thin films changed the sign, reflecting the anisotropic crystallographic orientations. For an ideal isotropic material, the CD taken through front and back illuminations were expected to be identical. The flipping of the peaks' signs suggested in the strong directional dependence of the signals along with strong LD and LB influences. As we formerly discussed, such anisropy in the material geometry introduces the LDLB artifacts. We quantified the actual CD and the LDLB artifacts by utilizing equations (1) and (2) respectively. We measured the actual CD spectra for the spin-coated thin films (Fig 4c). After averaging the front and back CD measurements, we observed the CD signals for R-Cl and S-Cl thin films as mirror images to each other in the band-absorption regime ~ 400 nm corresponding to intrinsic CD of the material. Here in accordance with the generic definition of CD, the CD of R-Cl showed negative feature while that of S-Cl showed positive feature. Notably, the intensity of the CD signals reduced by approximately 3 times. This means that the LD and LDLB effects below the absorption edge were suppressed. However, the material's intrinsic CD is firmly smeared out by the anisotropy related LDLB features. Next, we plotted the LDLB spectra for the spin-coated thin films (Fig 4d). We observed the spectral features in ~ 400 nm (same as CD position) whose intensity corresponded to what was missing in the actual CD spectra. The crystallographic misalignment resulted in the LDLB spectra for both R-Cl and S-Cl to show the same peak sign. Despite the suppression of the LD and LDLB effects, these artifacts persistently entangle with the material's intrinsic CD. This was validated by the fact that spectral features for the CD and LDLB overlap causing the misjudgment in the actual CD and the LDLB artifacts.

In contrast, the slow diffused thin films exhibited noticeably different chiroptical features. As illustrated in figure 4a (right), the $CD_{Front}$ signals shifted below the absorption edge, ∼ 430 nm. We also observed the broadening of the peak, extending approximately from 350 nm to 525 nm, comprising both band-absorption and below the absorption edges. The $CD_{Front}$ features for R-Cl and S-Cl thin films exhibited definite mirror-imaging. The S-Cl and R-Cl thin films showed positive and negative $CD_{Front}$ signs respectively, while the magnitude remained comparable. The shift in the peak positions may already indicate in the $CD_{Front}$ features being counterfeit and highly influenced by the LD and LDLB effects. It is important to acknowledge that the slow-diffused thin films possess strong crystallographic orientations and larger crystal domains. This makes these thin films more viable towards directional artifacts. Next, we performed the $CD_{Back}$ measurement. As expected, the robust crystal orientation anisotropy resulted in the flipping in the signs of the $CD_{Back}$ signals for the R-Cl to positive and the S-Cl to negative (Fig 4b). The $CD_{Back}$ retained the similar peak positions (∼ 430 nm) and mirror-images like features. Then, we did the semi-sum of the $CD_{Front}$ and the $CD_{Back}$ to plot the actual CD spectra (Fig 4c). We noticed that the actual CD of the slow diffused samples was no longer mirror images of each other as the LD and LDLB effects were not fully eliminated. The CD signals for both R-Cl and S-Cl were observed inverted in ∼ 430 nm and the intensity reduced nealy 7 times compared to either the $CD_{Front}$ or the $CD_{Back}$. The reduction CD amplitude and the inconsistency in the CD signals with no mirror-imaging together confirm the significant contribution from LDLB. This outcome is in agreement with previously reported correction where it is expected in the lowering of CD responses which could be strongly dominated by LDLB[54,57].

Next, we plotted the LDLB contribution (Fig 4d) by applying the method suggested in equation (2). We observed pronounced peaks for R-Cl and S-Cl thin films located at ∼430 nm, validating strong LDLB effect due to oriented crystallites. The 430 nm peaks were entwined to the peaks observed in the CD spectra of slow diffused thin films in Figure 4(a-b, right). This implies that the observed CD in the slow-diffused thin films are heavily affected by the LBLD effects. When ignored, this undermines the overall validation of the experiment. Therefore, it is essential to distinguish between actual and false CD responses in chiral perovskites.

When we followed the approach shown by equation (1), it is interesting to note that the effects from LDLB are reduced in the actual CD spectra. This certain reduction is not enough to visually realize the actual CD in the slow-diffused thin films due to the lack of the mirror image line shape. Nevertheless, the actual CD signals that are anticipated at ∼380 - 400 nm regions are suppressed, indicating that the thin films have large crystallites and considerable macroscopic anisotropy. In Figure 4a, the higher value of CD in slow-diffused thin films (∼5 times greater than the CD in the spin coated thin films) makes it credible that controlled-oriented thin films exhibit better performance than popular spin-coated thin films. While it is challenging to precisely control the crystal orientation using the slow diffusion method, i.e., we obtained multiple branches of crystal grains, the highly oriented film does exhibit stronger and more distinct intrinsic CD than spin coated. This makes slow diffused thin films with desired orientation advantageous over spin coated counterparts in terms of higher CD and hence, potentially highly applicable in CPL sources and detectors.

In summary, these findings collectively demonstrate that the crystal orientations have a significant impact in the acquired CD signals in both spin coated and slow diffused samples. They depend on the delicate balance between extrinsic LDLB effects and intrinsic chiroptical response. The high degree of crystal orientation can potentially improve the CD in the chiral perovskite materials, the LDLB effects get enhanced with controlled orientation. It is thus important to develop a new method to tackle the concern of separating the LDLB artifacts from the observed CD while balancing the true CD signal in the chiral thin films.

## 2.5. Discrete σ⁺/σ⁻ Generation for Artifact-Free CD Measurements

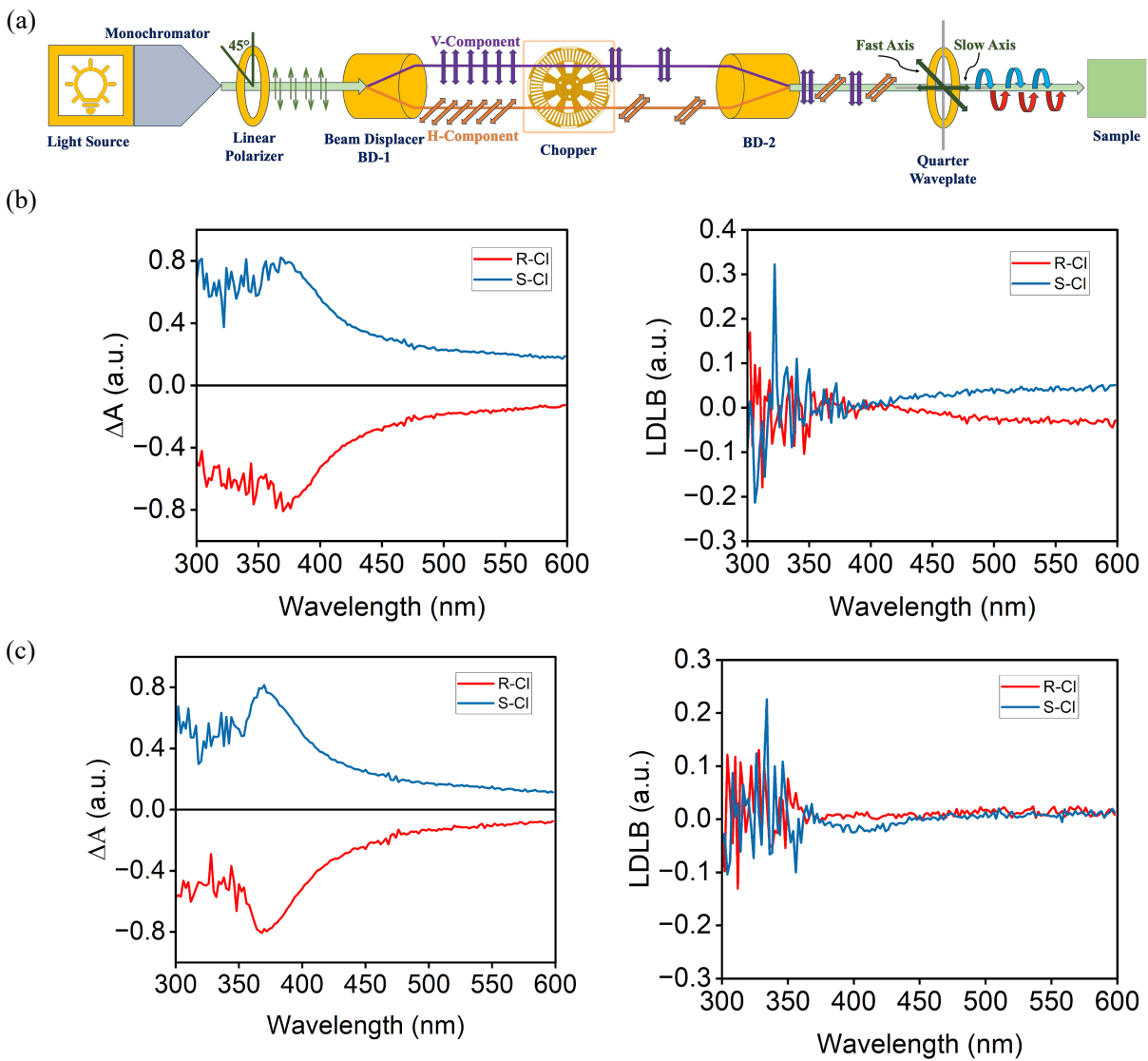

**Figure 5. Discrete circularly polarized light leading to modified CD signals and eliminated LDLB artifacts in thin films.** (a) Schematic illustration of CD experimental setup for the discrete generation of σ⁺ and σ⁻. The differential absorption (ΔA) spectra and LDLB spectra with discrete CPL (b) spin coated and (c) slow diffused thin films.

At first, in the conventional CD experimental system used for collecting CD in Figure 3-4, we generated the σ⁺ and σ⁻ circularly polarized light by utilizing a photo-elastic modulator (PEM, see

SI figure 06). In this setup, PEM sweep the polarization states continuously between the $\sigma^+$ and $\sigma^-$ with linear and elliptical signals present. This can introduce artifacts such as LD and LDLB during CD signal detection as demonstrated in our results. Photo-selection, emission from anisotropic samples, back-reflections from certain linear polarizers, and reflections or scatter of polarized excitation light are some of the sources of linear polarization[62]. In orientated samples, these remaining linear polarization components combine with anisotropic absorption or scattering to produce LDLB artifacts. The LDLB response contaminates the actual CD intrinsic to the sample and produce false CD response. Therefore, it is important to generate pure circularly polarized light free from linear polarization components.

Our hypothesis is that by excluding the intermediate polarization phases, pure circularly polarized light can be produced. This is in accordance with experimental planning aimed to generate pure circularly polarized light free from unfavorable influences from linearly polarized light components. For this, we adapted an experimental set-up proposed in the work done by Narushima et al.[56], which carefully eliminate linearly polarized light components during the generation of circularly polarized light. Any linear artifacts, including LDLB effects, are eliminated when there are no linearly polarized light components in the incident light (i.e. pure circularly polarized light incidence) on the samples.

The schematic illustration of the experimental set-up for the discrete generation of circularly polarized light is shown in the figure 5a. Thus, generated $\sigma^+$ and $\sigma^-$ are illuminated on samples. With this approach, we achieved the careful measurement of CD signals for both spin coated and slow diffused chiral thin films. Two spatially separated (4 mm) parallel, mutually orthogonal pristine linear polarization components were produced by the beam displacer BD-1. One of the divided two polarization components was always blocked by the chopper, and the subsequent optical system received alternating beams that were polarized horizontally and vertically. The two chopped beams were then resembled by BD-2, which was composed of coaxially coupled vertically and horizontally polarized beams. After passing these beams through the quarter wave-plate, the sample was illuminated with alternating circularly polarized light that was both $\sigma^+$ and $\sigma^-$. We detected the difference in absorption between $\sigma^+$ and $\sigma^-$ by the photo-detector connected with a lock-in amplifier.

The discrete generation of $\sigma^+$ and $\sigma^-$ favored in eliminating possible presence of linearly polarized light components that could contribute to annihilate the LDLB artifacts. The differential absorption ($\Delta A$) spectra along with the quantification of LDLB effects in both spin coated and slow diffused thin films following the schematic shown in figure 5a are shown in figure 5(b-c) respectively. The $\Delta A$ spectra clearly exhibited significant signals around 380 nm with no additional features below the absorption edge. This showed absence of any linear polarization components possibly responsible for LDLB artifacts. In addition to that, the LDLB spectra are displayed in Figure 5 b-c, which were quantified by the similar approach using equation (2). The absence of distinct peaks below the absorption edge ($\sim$ 430 nm) in the LDLB plots for both spin-coated and the slow-diffused thin films further supported the elimination of linear polarization related artifacts. To confirm our measurement, we also measured the CD for the spin-coated thin films with high throughput commercial CD System (see SI figure 07). The measured $\Delta A$ spectra for the thin films were in alignment with the spectra obtained from the commercial CD system. In both cases, the peak corresponding to the chiroptical response were observed $\sim$ 380 nm. The CD as high as $\sim$

1500 mdeg were observed in the both R-Cl and S-Cl spin coated thin films. This result was in alignment with the previously reported CD value[43]. However, such a large CD value possibly could have the contribution from the symmetric part of LDLB[54,57]. With the beam displacer integrated experimental arrangement, we were successfully able to optimize pure circularly polarized light. To check the reproducibility of the ΔA spectra, we also analyzed the ΔA with the front and back surfaces measurements, which produced the similar ΔA responses on both sides (SI Figure 08). Thus, this process ensured quantitative dependability for further analysis by verifying that the observed signal was coming exclusively from the inherent chiroptical response of the thin films.

## 3. Discussion and conclusion

In our work, we have developed a home-built CD experimental system which provides a wide range of customization add-ons allowing one to conduct multiple experiments. We studied chiroptical properties of (R-/S-MBA)$_2$CuCl$_4$ thin films focused on elucidating CD signal influenced by LD and LDLB artifacts as a result of crystal orientation. To understand the impact of the crystal orientation to the CD signal, we created oriented chiral perovskite thin films by slow diffusion method and compared with randomly oriented thin films fabricated with spin coating method. Both spin-coated and slow diffused thin films exhibited significant crystallites oriented in certain ways that favored substantial LD effect because of their anisotropy. On top of LD, the LB effect is caused by linearly polarized light being broken into two orthogonal components when passing through certain crystal plane of the material. The LDLB effects combined add up to the observed CD due to chiral material's handedness and selective properties. Thus, caution must be taken to eliminate the LD, LDLB effects to obtain the real CD response in highly crystalline samples. We developed a new optical path by removing the intermediate phases of the polarization states, yielding minimal contribution from the LDLB effects. Our CD results showed significant improvement with elimination of LDLB effects. We verified the reproducibility of the results and succeeded. Thus, we presented a thorough experimental work thereby assuring empirical consistency for additional study.

## 4. Method

### 4.1. Reagents and Materials

Chemicals including (R)-(+)-α-Methylbenzylamine (R-MBA, 98%), (S)-(-)-α-Methylbenzylamine (S-MBA, 98%) and 47% aqueous hydrochloric acid (HCl) solution were purchased from Sigma-Aldrich. Dimethylformamide (DMF, > 99.7%), and chlorobenzene (CB, > 99.0%) were purchased from Thermo Fisher Scientific (Fisher Bioreagents). Absolute ethanol (200 proof), and ethyl ether (laboratory grade) were purchased from Thermo Fisher Scientific (Fisher Chemicals). Isopropanol (>99.6%) was purchased from Thermo scientific. These chemical products were all utilized straight away, without any additional purification.

### 4.2. Synthesis of Chiral Ammonium Chloride Salts

Chiral-chloride based salts were prepared by adding 18 mmol (2.3 mL) of R-MBA (or, S-MBA) into 10 mL absolute ethanol. The solution was kept in ice-bath and when the temperature dropped down to 0°C, 19.5mmol of concentrated HCl aqueous solution was added to the solution under constant stirring in an ice water bath. The solution was kept at 0°C overnight and the overall solution was then heated at 80°C to get rid of excessive solvents until the white precipitates were observed. The precipitates were then treated multiple times with isopropanol and then with ether to get thick white paste which was then filtered and dried at 80°C to get white solid chiral ammonium chloride salt.

### 4.3. (R-/S-MBA)$_2$CuCl$_4$ Thin Film Fabrication Methods

Chiral thin films were fabricated by utilizing two approaches: spin-coating, which results in randomly oriented polycrystalline thin films, and slow-diffusion, resulting solvent gradient which yields well-defined directionally oriented layered crystals. The prepared ammonium chloride salts (R-/S-MBACl) and copper chloride (CuCl$_2$) were taken in the stoichiometric ratio of 2:1 and dissolved in 1 mL DMF to form 2D (R-/S-MBA)$_2$CuCl$_4$ precursors maintaining the molar ratio of 2 M for spin coating and 0.5 M for slow-diffusion methods. Thin films were fabricated on the 2 × 2 glass substrates. The glass substrates were ultrasonicated with ethanol for 15 minutes. They were dried and then ozone treated to enhance wettability with Ossila UV ozone cleaner for 20 minutes. For the spin coating method, the formulated precursor (2M concentration) was dropped perpendicularly without contact with the help of a Fisherbrand pipette on top of the glass substrate (we used 100μL), which was positioned in the spin coater. The spin coating was completed in the three steps: dropping, spinning and annealing. The spinning was concluded combining 1000 rpm for 10 seconds and 4000 rpm for 60 seconds. During the later spinning, 100μL chlorobenzene as an anti-solvent was dropped when 30 seconds was reached to assist on solidification of precursor. Next, the thin films were annealed at 60°C for 5 minutes followed by an extra layer of protection with 20mg PMMA in 1000μL chlorobenzene solution (100μL in each sample). For slow diffusion, the formulated precursor (0.5M concentration) was dropped (5μL on each sample) on the edge of the substrate and another substrate were used perpendicularly to drag the precursor across the bottom substrate to create solution gradient, and the system was left to grow controlled crystallites under the cover of crystallization dish to prevent air flow for 2 days.

### 4.4. Circular Dichroism and Linear Dichroism Measurements

The procedure for the CD measurement by photo-elastic modulator scheme has been thoroughly described in the supporting information (description of SI Figure 06). In the same set-up, LD measurement can be performed (description of SI Figure 07). On the other hand, for the CD measurement involving the beam displacer schematic (Figure 4a), we used the Spectral Products ASB-XE-E180 with 175W compact high-intensity xenon lamp (dominant broadband output 250 - 1100 nm) as light source, coupled to the Spectral Products DK40 monochromator with motorized slits (width: 10 μm - 3000 μm, height: 2 mm - 20 mm) to get the wavelength specific monochromate light. To generate linearly polarized light, Thorlabs Glan-Taylor GT10-A calcite linear polarizer was utilized, whose transmission axis was set at 45° relative to our lab reference frame, matching the fast axis orientation of Thorlabs AQWP05M-600 (ARC: 400-800nm) quarter wave-plate for discrete modulation of circularly polarized light between $\sigma^+$ and $\sigma^-$. To optimize the purity of circularly polarized light, the quality of linearly polarized light should also be considered. This criterion was met by incorporating Thorlabs BD40 - Calcite beam displacers (BD-1 and BD-2) in the experimental arrangement. Here, BD-1 generated two spatially separated (4mm) parallel mutually orthogonal pristine linear polarization components (x and y polarized light). The two components were next modulated by Thorlabs MC1F10A 10 slot blade with adjustable duty cycle (0 - 50%) optical chopper controlled via MC2000B optical chopper controller system at 500 rpm. Then, BD-2 resembled the two chopped beams into a single beam that constituents coaxially combined horizontally and vertically polarized beams. We achieved $\sigma^+$ and $\sigma^-$ with quarter wave-plate. The Thorlabs DET10A/M Silicon detector (200–1100 nm), which was connected to the SR860 Lock-in Amplifier, was used to detect the light after it had passed through the sample. The lock-in amplifier detected the differential absorption between $\sigma^+$ and $\sigma^-$.

## 4.5. Characterization Methods

We performed crystal structure characterizations of thin films fabricated with spin coating and slow diffusion approaches by utilizing optical microscopy, scanning electron microscopy, and grazing incidence wide angle X-ray spectroscopy. For optical microscopy, we used AmScope ME520T trinocular compound microscope (lamp: 24V 100W) integrated with AmScope MU1000-HS microscope digital camera (10MP Aptina color CMOS) for imaging. The morphologies of thin films were examined by Focused Ion Beam Scanning Electron Microscope (FIB-SEM) - Carl Zeiss AURIGA CrossBeam at Science & Engineering Shared Facilities – University at Buffalo (UB) Cleanroom. GIWAXS data was collected on Xenocs Xeuss 3.0 X-ray scattering instrument at UB Department of Chemistry, Chemistry Instrumentation Center (CIC). Data was collected using the auxiliary Cu-source operating at 50 kV and 0.6 mA. LaB6 was used to calibrate the sample to detector distance (SDD) before each series of exposures. Sample alignment was performed via a series of z and incidence angle (α) scans. After alignment α was set to 0.15°, SDD was set to 80 mm, horizontal and vertical slits were set to 0.400 mm and 0.200 mm, respectively. Exposure times were 600 seconds. Wedge correction was performed using XACT.

## Acknowledgements

R.S. and W.N. acknowledge the research foundation of University at Buffalo for funding this work. R.S. and W.N. acknowledge the funding support from National Science Foundation DMR-2532768. We appreciate Dr. Matthew R. Crawley, University of Bufalo, Department of Chemistry for providing the training on GIWAXS along with his expertise for insightful analysis ideas. The author appreciates her colleagues Sharmistha Khan and Sailesh Bataju for their support during

**Conflicts of Interest**

The authors declare no conflict of interest.

**Data Availability Statement**

Upon reasonable request, the corresponding author will provide the data supporting the findings and conclusion of this study.

# Supplementary Information

*for*

**Impact of crystallinity on the circular and linear dichroism signals in chiral perovskite**

*Reshna Shrestha[1], Wanyi Nie[1*]*

**Affiliations:**

[1]Department of Physics, University at Buffalo, State University of New York, Buffalo, New York 14260, USA

*Corresponding to: wanyinie@buffalo.edu

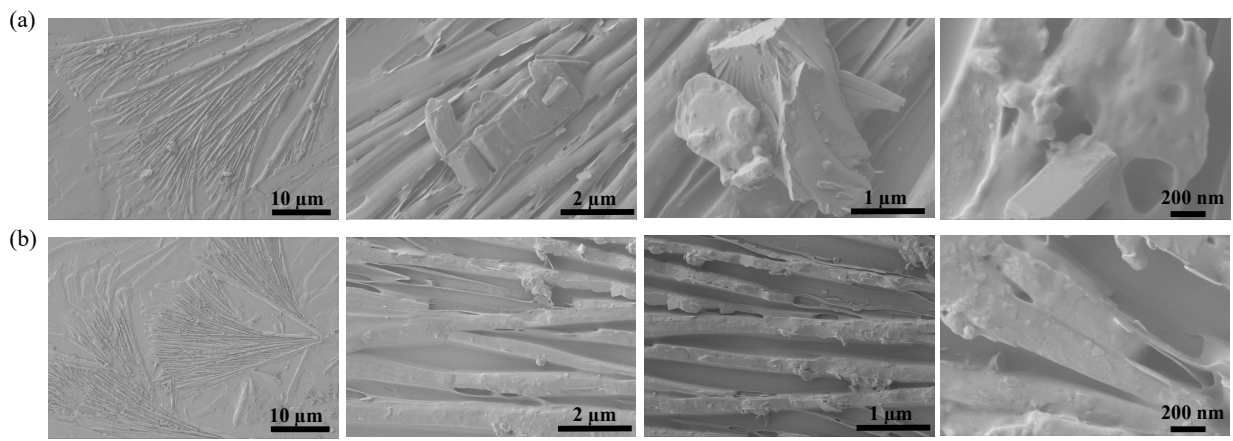

**SI Figure 01** SEM images of spin-coated (a) (R-MBA)$_2$CuCl$_4$ and (b) (S-MBA)$_2$CuCl$_4$ thin films.

The film morphology clearly showed micron-size random crystallite domains stacked together without any long-range ordering. Further zoom-in to nanoscale, both R-Cl and S-Cl exhibited rough surface morphology.

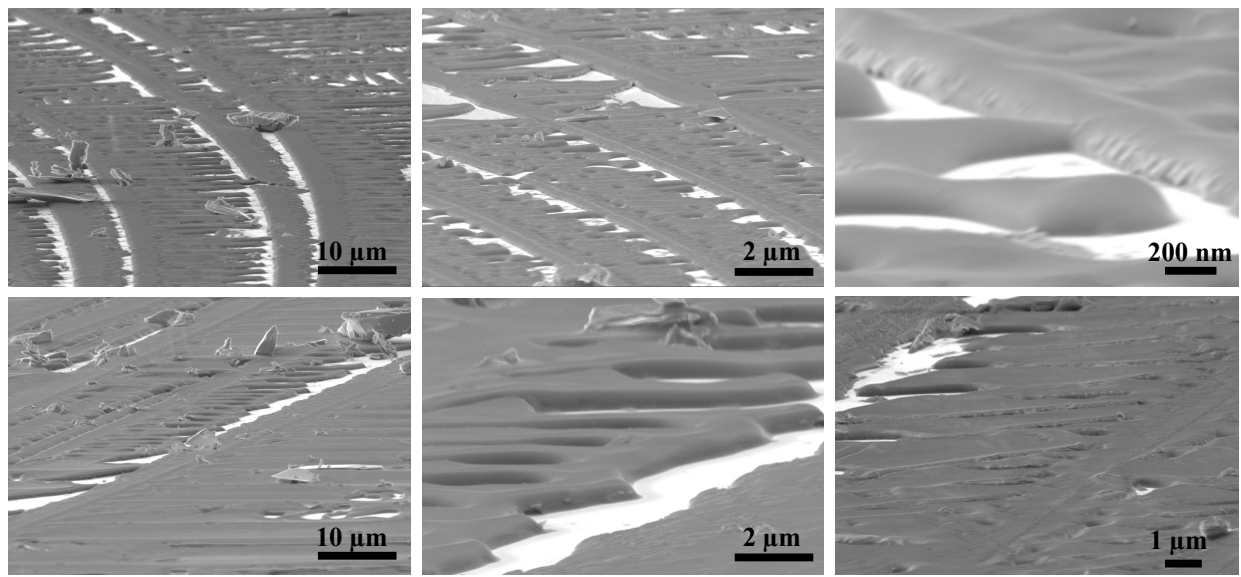

**SI Figure 02** SEM images of slow diffused (R-MBA)$_2$CuCl$_4$ thin films.

The cross-sectional observation of (R-MBA)$_2$CuCl$_4$ thin film provided better illustration of the crystal structure orientation. The images exhibited well-ordered layers of crystallites with even curved borders on micron and nano scales.

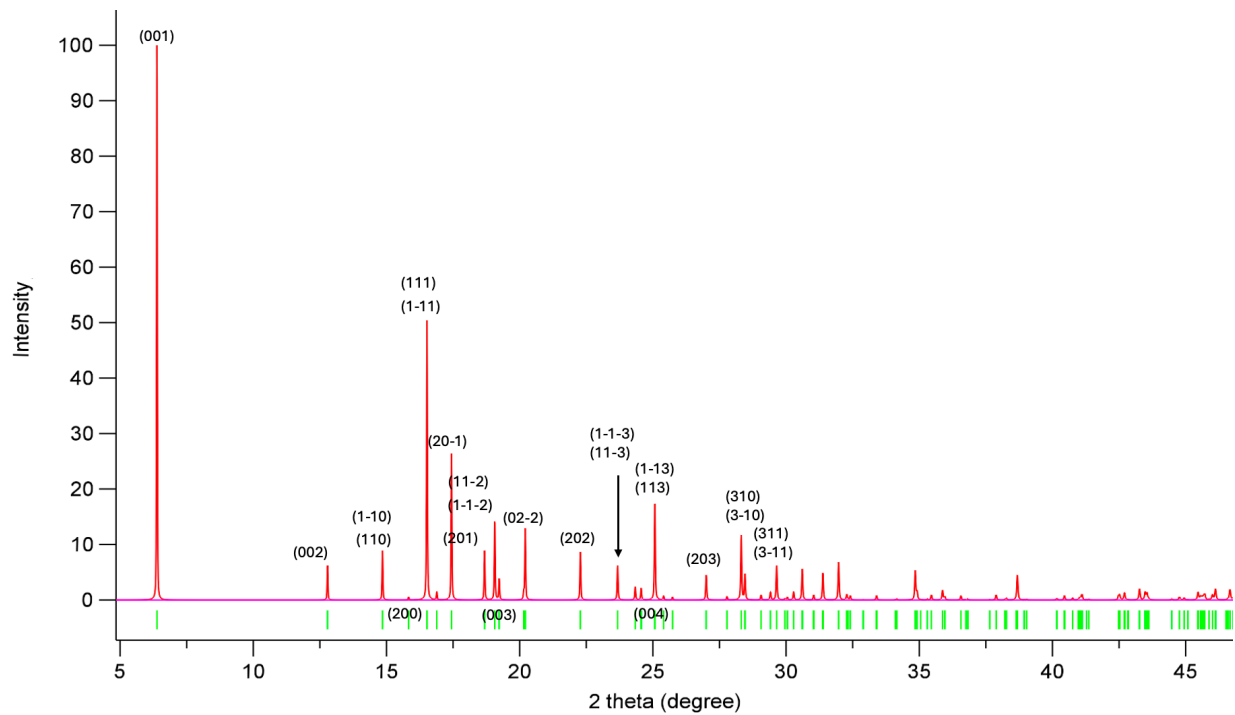

**SI Figure 03** Simulated powder X-ray diffraction (PXRD) patterns of (R/S-MBA)$_2$CuCl$_4$.

The PXRD patterns are simulated by utilizing the CIF of (R/S-MBA)$_2$CuCl$_4$ single crystals from the previously reported work[1]. The diffractions peaks were labeled by matching the peak intensities at 2θ (Bragg's diffraction angles) with corresponding crystallographic (hkl) planes.

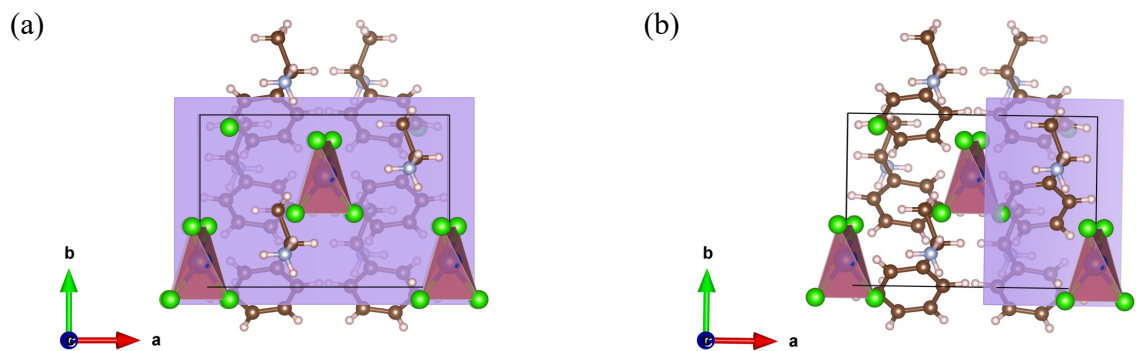

**SI Figure 04** Crystal structure visualization of (R-MBA)$_2$CuCl$_4$ in (a) (001) and (b) (201) crystallographic planes

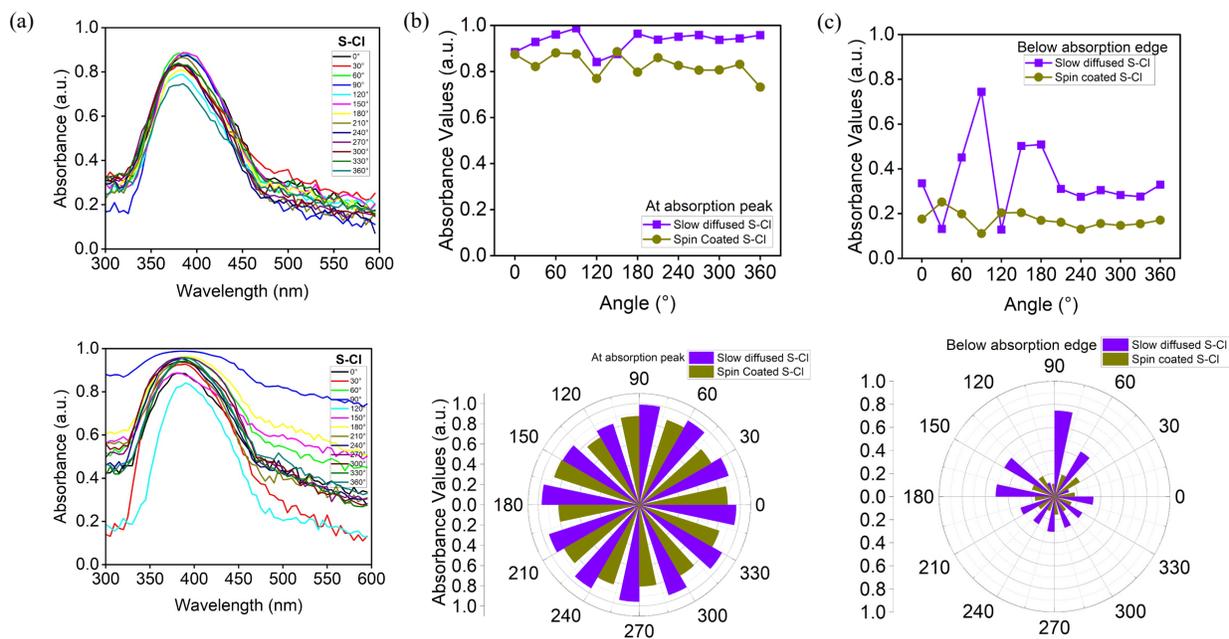

**SI Figure 05** Absorption of (S-MBA)$_2$CuCl$_4$ thin films taken with linearly polarized light with tuning polarizer angles. Absorption spectra of (a) Spin coated (top) and slow diffused (bottom) polycrystalline thin films. (b) Absorbance values at the absorption peak position (~ 380 nm) and (c) absorbance values below the absorption edge plotted as a function of LP orientation angle relative to the horizontal line in our lab reference frame.

The absorption trends for the (S-MBA)$_2$CuCl$_4$ thin films are approximately identical to that of (R-MBA)$_2$CuCl$_4$ thin films, as discussed in the main manuscript figure 2.

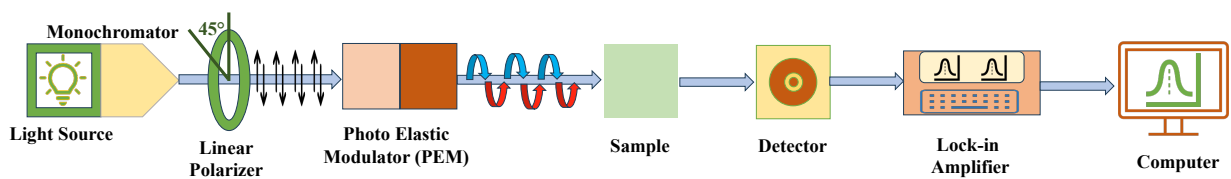

**SI Figure 06** Schematic illustration of the CD spectroscopy.

Here, we utilized the Spectral Products ASB-XE-E180 with 175W compact high-intensity xenon lamp (dominant broadband output 250 - 1100 nm) light source which was mono-chromated using the Spectral Products DK40 monochromator in the 300-600 nm wavelength range. The monochromated unpolarized light was linearly polarized by passing through the Thorlabs Glan-Taylor GT10-A calcite linear polarizer, which was then modulated between ($\sigma^+/\sigma^-$) left and right-handed circularly polarized light by one-quarter of the wavelength retardation by means of Hinds Instruments photo-elastic modulator (PEM) controlled with Hinds PEM-CSC controller. The linear polarizer was set at 45° to the PEM axis so that light contained equal x and y components. The PEM then modulated the relative phase, creating $\sigma^+$ and $\sigma^-$ or modulated ellipticity. Without the 45° alignment, no $\sigma^+$ and $\sigma^-$ were generated, leaving behind only intensity modulation or linear birefringence effects. The circularly polarized light when illuminated on the chiral sample, owing to the intrinsic chirality, there would be selective absorption of one of the handedness over the other. This differential absorption of $\sigma^+$ and $\sigma^-$ is termed as CD. After passing through the sample, the light was detected with the Thorlabs PDA45 – SiPM Amplified Detector (320 – 900 nm) which was connected to the SR860 Lock-in Amplifier. The lock-in amplifier measured the difference in $\sigma^+$ and $\sigma^-$. Thus, collected data was plotted as a function of light wavelength resulting in the CD spectrum. Besides, the regular CD spectroscopy, our CD system has the potential of user-choice customization as well as a wide range of property-characterization integration such as temperature controller, cryostat, external magnetic field, external electric field and so on.

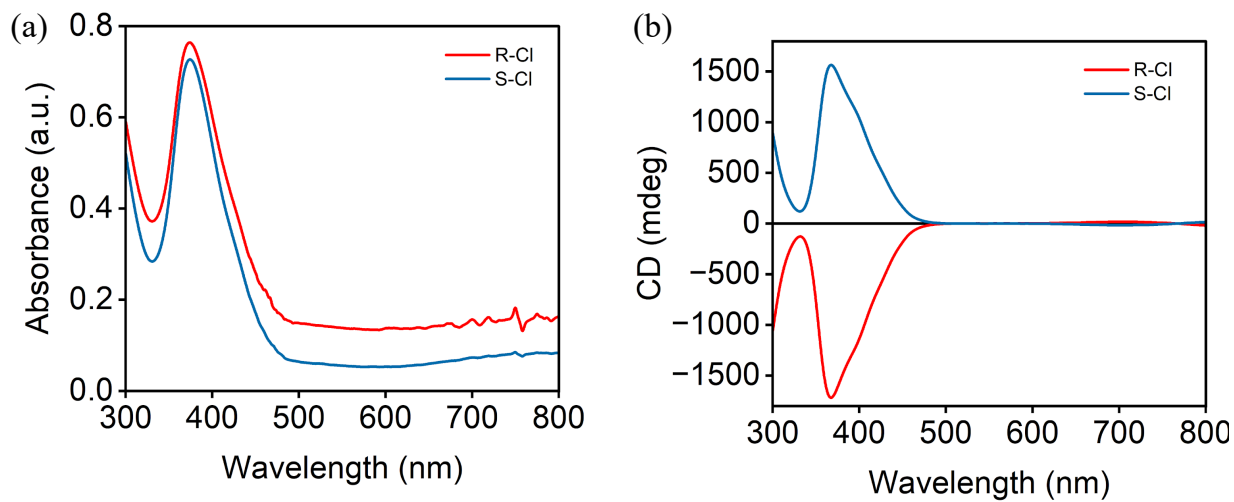

**SI Figure 07** (a) Absorption and (b) CD Spectra of spin coated (R/S-MBA)$_2$CuCl$_4$ thin films.

The CD signals for both (R-MBA)$_2$CuCl$_4$ and (S-MBA)$_2$CuCl$_4$ spin coated thin films were observed at ~380 nm corresponding to the excitation absorption peaks. The CD spectra for chiral right and left-handedness perovskites appeared exact mirror images to each other. The high CD signals located at the absorption peak position were in agreement with the differential absorption responses collected with discrete modulation of circularly polarized light approach.

**Circular Dichroism Measurement**

The CD measurement was carried out at Lawrence Berkely National Laboratory (LBNL) Catalysis Laboratory, 20 Lewis Hall, University of California, Berkeley. Both the absorption and CD data were acquired via high-throughput commercial CD system [Jasco CD (J-815) spectrometer] in 300-800 nm spectral range.

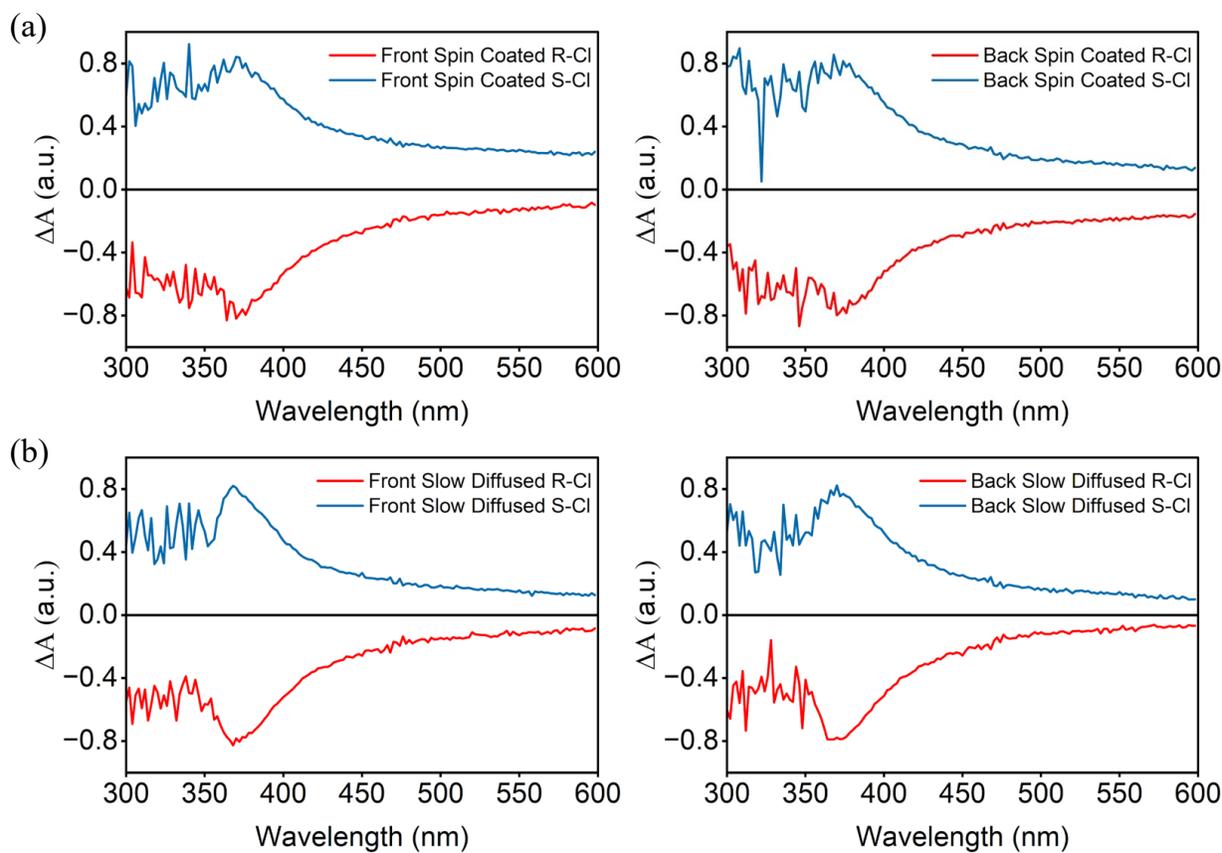

**SI Figure 08** Differential absorbance by front surface and back surface CD measurements of (a) spin-coated and (b) slow-diffused (R/S-MBA)$_2$CuCl$_4$ thin films.

We plotted the differential absorbance spectra for both spin-coated and slow-diffused thin films following the approach illustrated in the equation 1 of the main manuscript. The experimental arrangement was achieved by aligning the film facing the light source (front surface measurements) or the detector (back surface measurements). In both front and back measurements, the observed ΔA signals appeared identical with the peak response at ∼ 380 nm. With this, we confirmed the dependability and reproducibility of the experimental data in the beam displacers integrated CD system.